\documentclass{optica-article}

\journal{opticajournal} 

\articletype{Research Article}

\usepackage{graphicx} 
\usepackage{float} 
\usepackage{wrapfig} 
\usepackage[labelfont=bf]{caption} 
\usepackage{caption, subcaption} 

\usepackage{tikz,tikz-3dplot}
\usetikzlibrary{arrows.meta,bending,matrix,positioning,patterns,fit}

\usepackage{bm} 
\usepackage{hyperref} 
\usepackage{lineno} 

\begin{document}
\title{A mathematical model for inverse freeform design of a point-to-point two-reflector system}
\author{P. A. Braam,\authormark{1,*} J. H. M. ten Thije Boonkkamp,\authormark{1}\\ M. J. H. Anthonissen,\authormark{1} R. Beltman,\authormark{2} and W. L. IJzerman\authormark{1,2}}
\address{\authormark{1}CASA, Department of Mathematics and Computer Science, Eindhoven University of Technology, P.O. Box 513, 5600 MB Eindhoven, The Netherlands\\
\authormark{2}Signify Research, High Tech Campus 7, 5656 AE Eindhoven, The Netherlands}
\email{\authormark{*}p.a.braam@tue.nl}

\begin{abstract*}
In this paper, we discuss a mathematical model for inverse freeform design of an optical system with two reflectors in which light transfers from a point source to a point target. In this model, the angular light intensity emitted from the point source and illuminance arriving at the point target are specified by distributions. To determine the optical mapping and the shape of the reflectors, we use the optical path length and take energy conservation into account, through which we obtain a generated Jacobian equation. We express the system in both spherical and stereographic coordinates, and solve it using a sophisticated least-squares algorithm. Several examples illustrate the algorithm's capabilities to tackle complicated light distributions.
\end{abstract*}

\section{Introduction}
The goal in non-imaging optics is to design optical surfaces, such as reflectors and lenses, that convert a source light distribution to a target. To achieve this, many methodologies are used among which forward and inverse methods \cite{Luneburg,Optics_Eugene_Hecht}. Forward methods model the optical system through the random sampling of light rays, also known as ray-tracing and use the law of reflection and Snell's law \cite{RayT,Bosel1,Bosel2,Bosel3}. However, in complex optical systems, these methods often require manual adjustments of system parameters and rely on iterative techniques to gradually improve performance, whereas inverse methods use principles of geometrical optics and conservation of energy to compute the shapes of optical surfaces directly.

One approach to compute freeform surfaces in optical design is to solve a partial differential equation taking the form of a standard Monge-Ampère equation, a generalized Monge-Ampère equation or a generated Jacobian equation. We give a brief summary of literature on this approach; for a detailed overview, see Romijn \textit{et al.} \cite{Lotte}. To solve the partial differential equation in this approach, various methods exist: Kawecki \textit{et al.} used a finite element method \cite{Kawecki}, Wu \textit{et al.} used Newton's method \cite{Wu} and Brix \textit{et al.} used a collocation method with iteration techniques on a nonlinear solver \cite{Brix1,Brix2}. Moreover, Caboussat \textit{et al.} introduced a least-squares method \cite{LSmethod}, that was used by Prins \textit{et al.} to compute optical systems with one freeform surface \cite{Corien}. This method was first extended by Yadav \textit{et al.} to double freeform systems \cite{Nitin}, then by Romijn \textit{et al.} by taking generating functions as input \cite{Lotte} and later by van Roosmalen \textit{et al.} by also taking Fresnel reflections into account \cite{Teun1}.

In non-imaging systems we can distinguish 16 base optical systems that can be combined to create more complex optical configurations \cite{Martijn1}. These systems have two types of sources: sources where light originates from a parallel beam and point sources. The light then traverses through either a reflector or a lens system. Finally, it reaches a desired target domain, which is located in one of four target areas: a region reached by a parallel beam, a point target, a near-field target or a far-field target. The latter two areas indicate that the target area is located relatively close or far from the optical surface, compared to the size of the optical system.

\clearpage

\begin{figure}[!ht]
  \centering
  
  \begin{subfigure}[b]{0.58\textwidth}
      \centering
      \includegraphics[width=1\textwidth,trim={5cm 18.6cm 10.5cm 4.5cm},clip]{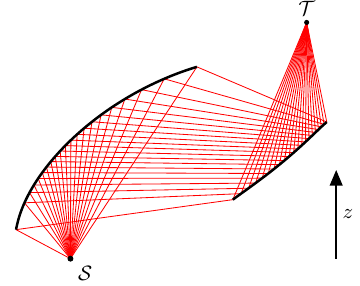}
  \end{subfigure}
  \hfill
  \begin{subfigure}[b]{0.38\textwidth}
      \centering
      \includegraphics[width=1\textwidth,trim={5cm 8.7cm 12cm 17cm},clip]{Figures/figure1.pdf}
      
      \vspace{0.5cm} 
      \includegraphics[width=1\textwidth,trim={5cm 11.6cm 12cm 14cm},clip]{Figures/figure1.pdf}
  \end{subfigure}
  
  \caption{A point-to-point two-reflector system with a normal light intensity $I_{\mathcal{S}}$ at the source $\mathcal{S}$ and a uniform light intensity $I_{\mathcal{T}}$ at the target $\mathcal{T}$. Light ray angles $\phi$ at the source and $\psi$ at the target are with respect to the optical $z-$axis.}
  \label{fig:General_point_to_point_reflector_system}
\end{figure}

This paper focuses on a specific type of optical system: the point-to-point two-reflector system, which is illustrated in Fig. \ref{fig:General_point_to_point_reflector_system}. In this system, we are interested in finding the shape of the reflectors that transfer light emitted from a point source with a specific angular light distribution to a point target with a desired angular light distribution. Fig. \ref{fig:General_point_to_point_reflector_system} shows this system for a normal distribution at the point source and a uniform distribution at the point target. Notably, two reflectors are necessary to both redirect the light rays to the point target and meet the required light distribution in this point. This specific system can for instance be used in single-mode fiber optics to convert the light distribution at a source point to a different one using reflectors. We compute the reflectors in the system by applying a two-stage least-squares algorithm. In the first stage, the algorithm computes the optical ray mapping, defining where each light ray originating from the source will end up at the target, and in the second stage the shapes of the reflectors are determined.

The remainder of this paper is structured as follows. In Section \ref{sec:Formulation} we describe the point-to-point two-reflector system and introduce stereographic coordinates. Moreover, we use the optical path length and energy conservation to find the system's optical mapping. In Section \ref{sec:algorithm} we will provide an overview of the least-squares algorithm and elaborate on the computation of the reflector surfaces. Numerical results are obtained and discussed in Section \ref{sec:results}. Finally, in Section \ref{sec:conclusion} we summarize the conclusions and propose recommendations for further research.

\section{Formulation of the point-to-point two-reflector system}\label{sec:Formulation}
In this section, we will first formulate a mathematical model for the point-to-point two-reflector system using the optical path length formulated in terms of stereographic coordinates. By using energy conservation and convexity, we then show how the system's optical mapping and reflector surfaces can be obtained. 

In the remainder of this paper, we denote vectors of unit length with a hat, e.g., $\hat{\bm{s}}$.

\subsection{Stereographic coordinates}
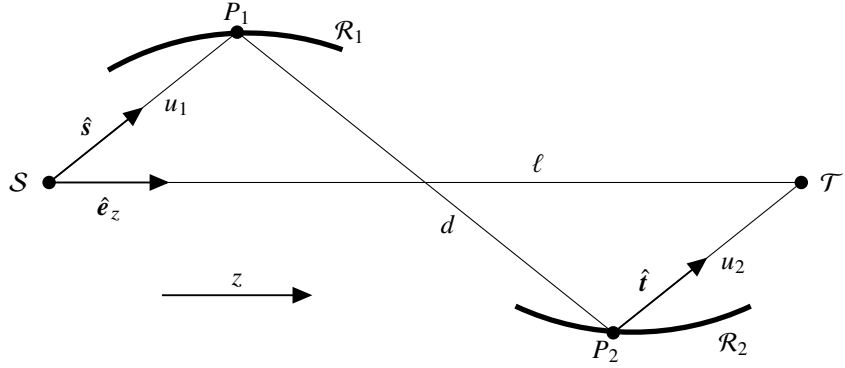
\begin{figure}[!ht]
\begin{center}
\begin{tikzpicture}[scale=1.0]
    \draw (0,0) -- (10,0);
    \draw (0,0) -- (2.5,2);
    \draw (2.5,2) -- (7.5,-2);
    \draw (7.5,-2) -- (10,0);
    \node at (0,0)[circle,fill,inner sep=1.8pt]{};
    \node at (2.5,2)[circle,fill,inner sep=1.8pt]{};
    \node at (7.5,-2)[circle,fill,inner sep=1.8pt]{};
    \node at (10,0)[circle,fill,inner sep=1.8pt]{};
    \node at (-0.4,0) {$\mathcal{S}$};
    \node at (10.4,0) {$\mathcal{T}$};
    \draw [line width=0.25mm, >=triangle 45, ->] (0,0) -- (1.25,1);
    \draw [line width=0.25mm, >=triangle 45, ->] (7.5,-2) -- (8.75,-1);
    \draw [line width=0.25mm, >=triangle 45, ->] (0,0) -- (1.6,0);
    \draw [line width=0.7mm] (3.9,1.766) arc (70:120:3.7cm);
    \draw [line width=0.7mm] (6.2,-1.645) arc (245:295:3.7cm);
    \node at (0.5,0.75) {$\hat{\bm{s}}$};
    \node at (7.9,-1.3) {$\hat{\bm{t}}$};
    \node at (0.8,-0.35) {$\hat{\bm{e}}_z$};
    \node at (4,2) {$\mathcal{R}_1$};
    \node at (9.1,-2.15) {$\mathcal{R}_2$};
    \node at (1.7,1) {$u_1$};
    \node at (5.3,-0.55) {$d$};
    \node at (9.1,-1.1) {$u_2$};
    \node at (6.5,0.25) {$\ell$};
    \node at (2.5,2.25) {$P_1$};
    \node at (7.4,-2.25) {$P_2$};
    \draw [line width=0.25mm, >=triangle 45, ->] (1.5,-1.5) -- (3.5,-1.5);
    \node at (2.5,-1.3) {$z$};
\end{tikzpicture}
\end{center}
\caption{A typical light ray in the point-to-point two-reflector system.}
\label{fig:Point_to_point_reflector_system}
\end{figure}

We consider the point-to-point two-reflector system depicted in Fig. \ref{fig:Point_to_point_reflector_system}. A light ray originates from a point source $\mathcal{S}$ with direction $\hat{\bm{s}}$ and hits a first reflector $\mathcal{R}_1$ in point $P_1$. The ray then reflects and hits a second reflector $\mathcal{R}_2$ in point $P_2$. Finally, it arrives at the point target $\mathcal{T}$ with direction $\hat{\bm{t}}$. In Fig. \ref{fig:Point_to_point_reflector_system}, $u_1=u_1(\hat{\bm{s}})$ is the distance from $\mathcal{S}$ to $P_1$, $d$ is the distance from $P_1$ to $P_2$ and $u_2=u_2(\hat{\bm{t}})$ is the distance from $P_2$ to $\mathcal{T}$. Moreover, $\hat{\bm{e}}_z=(0,0,1)$ and $\ell$ denotes the distance from $\mathcal{S}$ to $\mathcal{T}$. Reflectors $\mathcal{R}_1$ and $\mathcal{R}_2$ can then be parameterized as $\bm{r}_1(\hat{\bm{s}})=u_1(\hat{\bm{s}})\hat{\bm{s}}$ and $\bm{r}_2(\hat{\bm{t}})=\ell\hat{\bm{e}}_z-u_2(\hat{\bm{t}})\hat{\bm{t}}$.

\begin{figure}[!ht]
\centering
\begin{center}
\begin{tikzpicture}[scale=1.5,tdplot_main_coords]
    \fill[gray!100,opacity=0.08] (-3,3) -- (3,3) -- (3,-3) -- (-3,-3) -- (-3,-3);
    
    \tdplotsetrotatedcoords{20}{80}{70}
    \draw [ball color=black!8,thin,tdplot_rotated_coords] (0,0,0) circle (1);
    
    \begin{scope}
        \clip (0.65,-2) rectangle (-2,2);
        \draw [dashed, thick] (0,0,0) circle (0.98);
    \end{scope}

    \begin{scope}
        \clip (0.65,-2) rectangle (2,2);
        \draw [thick] (0,0) circle (0.98);
    \end{scope}

    \draw [thick] (-3,0.1) -- (-3,3) -- (3,3) -- (3,-3) -- (-3,-3) -- (-3,-1.65);
    \draw [dashed, thick] (-3,0.1) -- (-3,-1.65);

    \node at (3.9,-2.4) {$z=0$};

    \node at (5.6,1.5)[circle,fill,inner sep=1.5pt]{}; 
    \node at (4.56,0.7)[circle,fill,inner sep=1.5pt]{}; 
    \node at (1,-2)[circle,fill,inner sep=1.5pt]{}; 
    \node at (-4,-0.5)[circle,fill,inner sep=1.5pt]{}; 
    \node at (0,0.35)[circle,fill,inner sep=1.5pt]{}; 
    
    \node at (6.5,1.6) {$S$}; 
    \node at (5.6,0.8) {$Q_1$}; 
    \node at (1,-2.25) {$Q_1'$}; 
    \node at (-5.1,-0.55) {$Q_2$}; 
    \node at (0,0.15) {$Q_2'$}; 

    \draw [dashed, thick] (5.6,1.5) -- (4.56,0.7);
    \draw [thick] (4.56,0.7) -- (3,-0.5);
    \draw [dashed, thick] (3,-0.5) -- (1,-2);

    \draw [dashed, thick] (5.4,1.5) -- (-4,-0.5);
\end{tikzpicture}
\end{center}
\caption{Stereographic projections from the south pole $S$ where points $Q_1$ and $Q_2$ on the unit sphere are projected to points $Q_1'$ and $Q_2'$ on the plane $z=0$, respectively.}
\label{fig:Stereographic_Projection}
\end{figure}
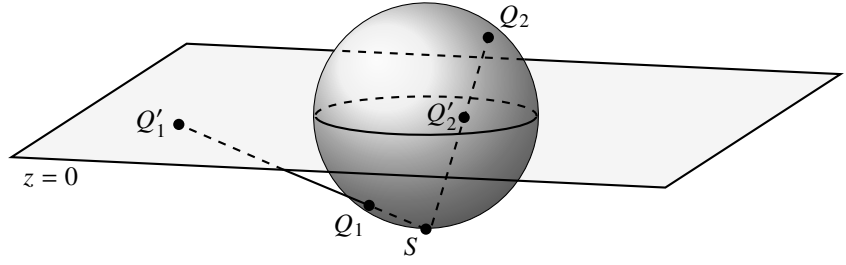

The source vector $\hat{\bm{s}}$ at the origin is parameterized by the zenith angle $\phi$ and the azimuth angle $\theta$ in spherical coordinates with $0\leq \phi\leq \pi$ and $0\leq \theta<2\pi$, and likewise the target vector $\hat{\bm{t}}$ is parameterized by the zenith angle $\psi$ and azimuth angle $\chi$ with $0\leq \psi\leq \pi$ and $0\leq \chi<2\pi$. We define
\begin{subequations}
\begin{align}\label{eq:s_and_t_trig}
&\hat{\bm{s}}=\begin{pmatrix}
    s_1\\[-3pt]
    s_2\\[-3pt]
    s_3
\end{pmatrix}
=
\begin{pmatrix}
    \sin(\phi)\cos(\theta)\\[-3pt]
    \sin(\phi)\sin(\theta)\\[-3pt]
    \cos(\phi)
\end{pmatrix},&&
\hat{\bm{t}}=\begin{pmatrix}
    t_1\\[-3pt]
    t_2\\[-3pt]
    t_3
\end{pmatrix}
=
\begin{pmatrix}
    \sin(\psi)\cos(\chi)\\[-3pt]
    \sin(\psi)\sin(\chi)\\[-3pt]
    \cos(\psi)
\end{pmatrix}.
\end{align}

To model the source and target vector, we use stereographic coordinates, which project the unit sphere onto the plane $z=0$. This system system can be used to describe the outgoing light in the point source and the incoming light in the point target. Therefore, we define stereographic projections with respect to the south pole as
\begin{align}\label{eq:stereographic_projections}
    &\bm{x}(\hat{\bm{s}})=\begin{pmatrix}x_1\\x_2\end{pmatrix}=\frac{1}{1+ s_3}\begin{pmatrix}
        s_1\\s_2
    \end{pmatrix},&&
    \bm{y}(\hat{\bm{t}})=\begin{pmatrix}y_1\\y_2\end{pmatrix}=\frac{1}{1+ t_3}\begin{pmatrix}
        t_1\\t_2
    \end{pmatrix},
\end{align}
which are undefined for $s_3=-1$ and $t_3=-1$ \cite{Lotte}. We take stereographic projections with respect to the south pole for both the source and point target, because this choice assures that a conical beam of rays directed in the positive $z-$direction yields a bounded source and target domain in stereographic coordinates, as illustrated in Fig. \ref{fig:Stereographic_Projection}. These projections have the corresponding inverse projections
\begin{align} \label{eq:inverse_stereographic_projections}
    &\hat{\bm{s}}(\bm{x})=\frac{1}{1+|\bm{x}|^2}\begin{pmatrix}
        2x_1\\[-3pt]2x_2\\[-3pt]1-|\bm{x}|^2
    \end{pmatrix},
    &&
    \hat{\bm{t}}(\bm{y})=\frac{1}{1+|\bm{y}|^2}\begin{pmatrix}
        2y_1\\[-3pt]2y_2\\[-3pt]1-|\bm{y}|^2
    \end{pmatrix}.
\end{align}
\end{subequations}

\tdplotsetmaincoords{80}{105}

\subsection{Optical path length}\label{sec:cost_function}
We aim to describe the reflector system using an equation of the form
\begin{align}\label{eq:transport_formulation}
    v_1(\hat{\bm{s}})+v_2(\hat{\bm{t}})=c(\hat{\bm{s}},\hat{\bm{t}}),
\end{align}
to describe how light rays from point source $\mathcal{S}$ end up at point target $\mathcal{T}$. In this equation, $v_1(\hat{\bm{s}})$ is a function related to the shape of the first reflector, $v_2(\hat{\bm{t}})$ to the second reflector and $c(\hat{\bm{s}},\hat{\bm{t}})$ is the cost function \cite{Villani,OptimalTransport}. To derive Eq. (\ref{eq:transport_formulation}), we introduce the optical path length
\begin{align}\label{eq:point_char_ptp}
    V=u_1(\hat{\bm{s}}) + d + u_2(\hat{\bm{t}}),
\end{align}
which is constant by the principle of equal optical path length, also known as the theorem of Malus and Dupin \cite{BornAndWolf}. Here, $d$ can be expressed as
\begin{align}\label{eq:dist_ptp}
    d&=|\ell\hat{\bm{e}}_z-u_1(\hat{\bm{s}})\hat{\bm{s}}-u_2(\hat{\bm{t}})\hat{\bm{t}}|.
\end{align}
By squaring Eq. (\ref{eq:point_char_ptp}), substituting Eq. (\ref{eq:dist_ptp}), and subsequently simplifying the resulting expression, we obtain
\begin{align}
    V^2=&\; u_1^2(\hat{\bm{s}})+d^2+u_2^2(\hat{\bm{t}})+2u_1(\hat{\bm{s}})u_2(\hat{\bm{t}})+2du_1({\hat{\bm{s}}})+2du_2(\hat{\bm{t}})\nonumber\\
    =&\;
    \ell^2+2u_1(\hat{\bm{s}})(V-\ell\hat{\bm{e}}_z\boldsymbol{\cdot}\hat{\bm{s}})+2u_2(\hat{\bm{t}})(V-\ell\hat{\bm{e}}_z\boldsymbol{\cdot}\hat{\bm{t}})-2u_1(\hat{\bm{s}})u_2(\hat{\bm{t}})(1-\hat{\bm{s}}\boldsymbol{\cdot}\hat{\bm{t}}),\nonumber
\end{align}
which can be factorized as
\begin{align}\label{eq:big_transport_equation_ptp2}
    \left(\frac{V^2-\ell^2}{2u_1(\hat{\bm{s}})(V-\ell\hat{\bm{e}}_z\boldsymbol{\cdot}\hat{\bm{s}})}-1\right)\left(\frac{V^2-\ell^2}{2u_2(\hat{\bm{t}})(V-\ell\hat{\bm{e}}_z\boldsymbol{\cdot}\hat{\bm{t}})}-1\right)=1-\frac{V^2-\ell^2}{2}\cdot\frac{1-\hat{\bm{s}}\boldsymbol{\cdot}\hat{\bm{t}}}{(V-\ell\hat{\bm{e}}_z\boldsymbol{\cdot}\hat{\bm{s}})(V-\ell\hat{\bm{e}}_z\boldsymbol{\cdot}\hat{\bm{t}})}.
\end{align}
By taking the logarithm on both sides, we obtain Eq. (\ref{eq:transport_formulation}) with
\begin{subequations}\label{eq:transport_old}
\begin{align}
    v_1(\hat{\bm{s}})&= \log\left(\frac{V^2-\ell^2}{2u_1(\hat{\bm{s}})(V-\ell\hat{\bm{e}}_z\boldsymbol{\cdot}\hat{\bm{s}})}-1\right),\label{eq:transport_u1_ptp}\\
    v_2(\hat{\bm{t}})&= \log\left(\frac{V^2-\ell^2}{2u_2(\hat{\bm{t}})(V-\ell\hat{\bm{e}}_z\boldsymbol{\cdot}\hat{\bm{t}})}-1\right),\label{eq:transport_u2_ptp}\\
    c(\hat{\bm{s}},\hat{\bm{t}})&=\log\left(1-\frac{V^2-\ell^2}{2}\cdot\frac{1-\hat{\bm{s}}\boldsymbol{\cdot}\hat{\bm{t}}}{(V-\ell\hat{\bm{e}}_z\boldsymbol{\cdot}\hat{\bm{s}})(V-\ell\hat{\bm{e}}_z\boldsymbol{\cdot}\hat{\bm{t}})}\right).\label{eq:transport_c_ptp}
\end{align}
\end{subequations}
In Appendix \ref{sec:appendix_A} we show that the arguments of the logarithms are strictly positive.
\par
Next, we aim to rewrite Eq. (\ref{eq:transport_formulation}) in terms of stereographic coordinates. Using the expressions in Eq. (\ref{eq:inverse_stereographic_projections}), the inner product of $\hat{\bm{s}}$ and $\hat{\bm{t}}$ can be written as
\begin{align*}
    \hat{\bm{s}}\bm{\cdot}\hat{\bm{t}}&=
    1-\frac{2|\bm{x}-\bm{y}|^2}{(1+|\bm{x}|^2)(1+|\bm{y}|^2)}.
\end{align*}
Consequently, we can write Eqs. (\ref{eq:transport_old}) in their stereographic form
\begin{subequations}
\label{eq:stereog_transport_ptp_easy}
    \begin{align}
    v_1(\bm{x})&= \log\left(\frac{(V^2-\ell^2)(1+|\bm{x}|^2)}{2u_1(\bm{x})\kappa_1(\bm{x})}-1\right),\label{eq:stereog_transport_u1_ptp_easy}
    \\
    v_2(\bm{y})&= \log\left(\frac{(V^2-\ell^2)(1+|\bm{y}|^2)}{2u_2(\bm{y})\kappa_2(\bm{y})}-1\right),\label{eq:stereog_transport_u2_ptp_easy}
    \\
    c(\bm{x},\bm{y})&=\log\left(1-\frac{\kappa_0(\bm{x},\bm{y})}{\kappa_1(\bm{x}) \kappa_2(\bm{y})}\right),\label{eq:stereog_transport_C_ptp_easy}
\end{align}
with the auxiliary variables
\begin{align}
    \kappa_0(\bm{x},\bm{y})&=(V^2-\ell^2)|\bm{x}-\bm{y}|^2,\label{eq:kappa0_in_stereographic_easy}\\
    \kappa_1(\bm{x})&=V-\ell+(V-\ell)|\bm{x}|^2,\label{eq:kappa1_in_stereographic_easy}\\
    \kappa_2(\bm{y})&=V-\ell+(V-\ell)|\bm{y}|^2.\label{eq:kappa2_in_stereographic_easy}
\end{align}
\end{subequations}
Therefore, Eq. (\ref{eq:transport_formulation}) can be written in the form
\begin{align}\label{eq:transport_equation_stereographic}
    v_1(\bm{x})+v_2(\bm{y})=c(\bm{x},\bm{y}).
\end{align}

\subsection{Energy conservation} \label{sec:energy_conservation}
We will now derive the energy balance in terms of stereographic coordinates $\bm{x}$ and $\bm{y}=\bm{m}(\bm{x})$ \cite{Corien,Jan1}, where $\bm{m}=\bm{m}(\bm{x})$ is the mapping that specifies how each of the light rays at the point source $\mathcal{S}$ will end up at the point target $\mathcal{T}$. We assume that $\mathcal{S}$ emits light with distribution $f(\bm{x})$ for $\bm{x}$ in the stereographic source domain $\mathcal{X}$, and maps this light to $\mathcal{T}$ with distribution $g(\bm{y})$ for $\bm{y}$ in the stereographic target domain $\mathcal{Y}$ via the reflector system given in Fig. \ref{fig:Point_to_point_reflector_system}. Energy conservation then gives
\begin{align*}
    \iint_{\mathcal{A}}f(\bm{x})\;\mathrm{d}\mathcal{S}(\bm{x})=\iint_{\bm{m}(\mathcal{A})}g(\bm{y})\;\mathrm{d}\mathcal{S}(\bm{y}),
\end{align*}
for arbitrary $\mathcal{A}\subset \mathcal{X}$. We can write this equation as
\begin{align*}
    \iint_{\mathcal{A}}f(\bm{x})\left|\frac{\partial \hat{\bm{s}}}{\partial x_1}\boldsymbol{\times} \frac{\partial \hat{\bm{s}}}{\partial x_2}\right|\mathrm{d}\bm{x}=\iint_{\bm{m}(\mathcal{A})}g(\bm{y})\left|\frac{\partial \hat{\bm{t}}}{\partial y_1}\boldsymbol{\times}\frac{\partial \hat{\bm{t}}}{\partial y_2}\right|\mathrm{d}\bm{y}.
\end{align*}
Using Eq. (\ref{eq:inverse_stereographic_projections}), this simplifies to
\begin{align*}
    \iint_{\mathcal{A}}\frac{4f(\bm{x})}{(1+|\bm{x}|^2)^2}\;\mathrm{d}\bm{x}=\iint_{\bm{m}(\mathcal{A})}\frac{4g(\bm{y})}{(1+|\bm{y}|^2)^2}\;\mathrm{d}\bm{y},
\end{align*}
and substituting the mapping $\bm{y}=\bm{m}(\bm{x})$ on the right-hand side, we find
\begin{align*}
    \iint_{\mathcal{A}}\frac{4f(\bm{x})}{(1+|\bm{x}|^2)^2}\;\mathrm{d}\bm{x}&=\iint_{\mathcal{A}}\frac{4g(\bm{m}(\bm{x}))}{(1+|\bm{m}(\bm{x})|^2)^2}\cdot |\text{det}(\text{D}\bm{m}(\bm{x}))|\;\mathrm{d}\bm{x},
\end{align*}
where $\text{D}\bm{m}(\bm{x})$ denotes the Jacobian of mapping $\bm{m}$. Assuming $\text{det}(\text{D}\bm{m}(\bm{x}))>0$, we obtain the generated Jacobian equation
\begin{align}\label{eq:Jacobian_equation_ptp}
    \text{det}(\text{D}\bm{m}(\bm{x}))&=\frac{(1+|\bm{m}(\bm{x})|^2)^2}{(1+|\bm{x}|^2)^2}\cdot\frac{f(\bm{x})}{g(\bm{m}(\bm{x}))}=:F(\bm{x},\bm{m}(\bm{x})).
\end{align}
In addition to the energy balance, we impose the transport boundary condition
\begin{align}\label{eq:transport_boundary_condition_BC}
    \bm{m}(\partial \mathcal{X})=\partial \mathcal{Y},
\end{align}
which ensures that the boundary of the source domain is mapped to the boundary of the target domain, and therefore that all the light from the source will be transferred to the target \cite{Teun1}.

\subsection{Convex solution} \label{sec:convex_analysis}
Eq. (\ref{eq:transport_equation_stereographic}) gives a relation between the two reflectors and the mapping with infinitely many solutions for $v_1$ and $v_2$. We take a unique solution by enforcing $v_1$ and $v_2$ to form a $c$-convex or $c$-concave pair \cite{Nitin,Teun,Boyd}. In the $c$-convex solution pair, $v_1$ and $v_2$ are given by
\begin{align*}
    v_1(\bm{x})=\max_{\bm{y}\in\mathcal{Y}}(c(\bm{x},\bm{y})-v_2(\bm{y})),&&
    v_2(\bm{y})=\max_{\bm{x}\in\mathcal{X}}(c(\bm{x},\bm{y})-v_1(\bm{x})).
\end{align*}
Likewise, the $c$-concave solution pair $v_1$ and $v_2$ is defined by
\begin{align*}
    v_1(\bm{x})=\min_{\bm{y}\in\mathcal{Y}}(c(\bm{x},\bm{y})-v_2(\bm{y})),&&
    v_2(\bm{y})=\min_{\bm{x}\in\mathcal{X}}(c(\bm{x},\bm{y})-v_1(\bm{x})).
\end{align*}
In both cases, $c(\cdot,\bm{y})-v_1$ has a stationary point and therefore we require
\begin{align}\label{eq:gradient_transport_equation}
    \nabla_{\bm{x}}c(\bm{x},\bm{y})-\nabla v_1(\bm{x})=\bm{0}.
\end{align}
By the implicit function theorem, this equation provides a mapping $\bm{y}=\bm{m}(\bm{x})$ under the condition that the matrix
\begin{align*}
    \bm{C}=\bm{C}(\bm{x},\bm{y})=\text{D}_{\bm{xy}}c=\left(\frac{\partial^2 c(\bm{x},\bm{y})}{\partial x_i\,\partial y_j}\right),
\end{align*}
is invertible, which is true for Eq. (\ref{eq:stereog_transport_C_ptp_easy}). Next, we substitute $\bm{y}=\bm{m}(\bm{x})$ into Eq. (\ref{eq:gradient_transport_equation}) and take the derivative with respect to $\bm{x}$ to obtain
\begin{align}\label{eq:CDm_is_P}
    \bm{C}\text{D}\bm{m}=\text{D}^2v_1-\text{D}_{\bm{xx}}c=:\bm{P}.
\end{align}
Notably, for a $c-$convex pair we require that the matrix $\bm{P}$ is symmetric positive definite (SPD) \cite{Lotte} and by Eq. (\ref{eq:Jacobian_equation_ptp}) that $\bm{P}$ satisfies
\begin{align}\label{eq:detP_constraint}
    \text{det}(\bm{P}(\bm{x}))=F(\bm{x},\bm{m}(\bm{x}))\text{det}(\bm{C}(\bm{x},\bm{m}(\bm{x}))).
\end{align}
In summary, we have to solve Eq. (\ref{eq:CDm_is_P}) subject to the constraint (\ref{eq:detP_constraint}) and the transport boundary condition (\ref{eq:transport_boundary_condition_BC}) for the mapping $\bm{m}(\bm{x})$. Subsequently, we have to solve Eq. (\ref{eq:gradient_transport_equation}) for $v_1(\bm{x})$ and finally obtain $u_1(\bm{x})$ and $u_2(\bm{y})$ using Eqs. (\ref{eq:stereog_transport_u1_ptp_easy}) and (\ref{eq:stereog_transport_u2_ptp_easy}).

\section{The least-squares algorithm}\label{sec:algorithm}
In this section we will outline the least-squares algorithm to simulate the point-to-point two-reflector system. In the least-squares algorithm, Eq. (\ref{eq:CDm_is_P}) is solved by minimizing the functional
\begin{align}\label{eq:functional_J_I}
    J_I[\bm{m},\bm{P}]=\tfrac{1}{2}\iint_{\mathcal{X}}||\bm{C}\text{D}\bm{m}-\bm{P}||^2\;\mathrm{d}\bm{x}
\end{align}
over $\bm{P}$, where $\bm{m}$ is fixed and where $\|\cdot\|$ denotes the Frobenius norm. Additionally, the transport boundary condition (\ref{eq:transport_boundary_condition_BC}) is imposed by minimizing the functional
\begin{align}\label{eq:functional_J_B}
    J_B[\bm{m},\bm{b}]=\tfrac{1}{2}\oint_{\partial \mathcal{X}}|\bm{m}-\bm{b}|^2\;\mathrm{d}\bm{x}
\end{align}
over $\bm{b}:\partial \mathcal{X}\to\partial\mathcal{Y}$, where $\bm{m}$ is fixed. Finally, the least-squares algorithm calculates the mapping $\bm{y}=\bm{m}(\bm{x})$ by minimizing the functional
\begin{align}\label{eq:functional_J}
    J[\bm{m},\bm{P},\bm{b}]=\alpha J_I[\bm{m},\bm{P}]+(1-\alpha)J_B[\bm{m},\bm{b}],
\end{align}
over $\bm{m}$ where $\alpha\in(0,1)$ is a weighting factor. The algorithm approximates the mapping $\bm{m}$ by starting from an initial mapping $\bm{m}^0$, which maps the source domain $\mathcal{X}$ uniformly to the target domain $\mathcal{Y}$, and then iterates according to
\begin{subequations}
\label{eq:overview_iteration}
\begin{align}
    \bm{b}^{n+1}&=\underset{\bm{b}\in\mathcal{B}}{\mathrm{argmin}}\;J_B[\bm{m}^n,\bm{b}],
    \\
    \bm{P}^{n+1}&=\underset{\bm{P}\in\mathcal{P}}{\mathrm{argmin}}\;J_I[\bm{m}^n,\bm{P}],
    \\
    \bm{m}^{n+1}&=\underset{\bm{m}\in\mathcal{M}}{\mathrm{argmin}}\;J[\bm{m},\bm{P}^{n+1},\bm{b}^{n+1}],\label{eq:overview_iteration_M}
\end{align}
for $n=0,1,2,..$, where the function spaces are defined by
\begin{align}
    \mathcal{B}&=\{\bm{b}\in C^1(\partial \mathcal{X})^2\mid\bm{b}(\bm{x})\in\partial\mathcal{Y}\},\\
    \mathcal{P}&=\{\bm{P}\in C^1(\mathcal{X})^{2\times2}\mid \bm{P}\text{ is SPD, }\text{det}(\bm{P}(\bm{x}))=F(\bm{x},\bm{m}(\bm{x}))\text{det}(\bm{C}(\bm{x},\bm{m}(\bm{x})))\},\\
    \mathcal{M}&=C^2(\mathcal{X})^2.
\end{align}
\end{subequations}
A more in-depth analysis can be found in \cite{Lotte,Teun,Rene}.

Upon convergence of (\ref{eq:overview_iteration}), we compute $u_1$ and $u_2$ by substituting $\bm{m}$ in Eq. (\ref{eq:gradient_transport_equation}). A solution for $v_1$ can then be found by minimizing the functional
\begin{align}\label{eq:functional_I}
    I[v_1]=\tfrac{1}{2}\int_\mathcal{X}|\nabla v_1(\bm{x})-\nabla_{\bm{x}}c(\bm{x},\bm{m}(\bm{x}))|^2\;\mathrm{d}\bm{x}.
\end{align}
By setting the first variation of this functional equal to zero, using Gauss's Theorem and the Fundamental Lemma of Calculus of Variations, we obtain the Neumann boundary value problem
\begin{subequations}\label{eq:BVP_first_surface}
\begin{align}
    \Delta v_1(\bm{x})&=\nabla\bm{\cdot}\nabla_{\bm{x}}c(\bm{x},\bm{m}),&&\bm{x}\in\mathcal{X},\label{eq:BVP_first_surface_1}\\
    \nabla v_1(\bm{x})\bm{\cdot}\hat{\bm{n}}&=\nabla_{\bm{x}}c(\bm{x},\bm{m})\bm{\cdot}\hat{\bm{n}},&&\bm{x}\in\partial \mathcal{X},\label{eq:BVP_first_surface_2}
\end{align}
\end{subequations}
where $\bm{\hat{n}}$ is the unit outward normal on the boundary of the source domain. Note that $v_1$ is determined up to an additive constant. We discretize this system for $v_1$ on a grid by using a finite volume method resulting in the linear system $\bm{A}\bm{v}=\bm{b}$. Since the boundary value problem (\ref{eq:BVP_first_surface}) has multiple solutions, we enforce a unique solution by setting the average distance from $\mathcal{S}$ to $\mathcal{R}_1$ equal to some $h\in\mathbb{R}$\cite{Lotte}. In Sec. \ref{sec:fixing_dist} we show that we can fix this distance for a specific point instead. Finally, by rewriting Eq. (\ref{eq:stereog_transport_u1_ptp_easy}) in the form
\begin{align*}
    & u_1(\bm{x})=\frac{(V^2-\ell^2)(1+|\bm{x}|^2)}{2\kappa_{1}(\bm{x})(e^{v_{1}(\bm{x})}+1)},
\end{align*}
we obtain the distance from the point source $\mathcal{S}$ to the first reflector $\mathcal{R}_1$ at every grid point. Likewise, we have
\begin{align*}
    u_2(\bm{y})=\frac{(V^2-\ell^2)(1+|\bm{y}|^2)}{2\kappa_{2}(\bm{y})(e^{v_{2}(\bm{y})}+1)},
\end{align*}
where $v_2$ is obtained using Eq. (\ref{eq:transport_equation_stereographic}). Thus, for every ray of light with stereographic coordinates $\bm{x}$ we know its mapping $\bm{y}=\bm{m}(\bm{x})$, the distance $u_1$ to the first reflector and the distance $u_2$ between the second reflector and point target $\mathcal{T}$. Therefore, the shape of the reflectors can be calculated by using the inverse projections (\ref{eq:inverse_stereographic_projections}) and the parameterizations of the reflectors.

\section{Numerical results}\label{sec:results}
In this section we will analyze two numerical examples. We first consider an example which demonstrates how the distance to the first reflector can be fixed for a specific ray of light and then look at an example with a complicated target distribution, which we validate by ray-tracing.

\subsection{Fixing the distance to the first reflector}\label{sec:fixing_dist}
\begin{figure}[!ht]
  \centering
\begin{subfigure}[b]{0.43\textwidth}
    \includegraphics[width = 1.0\textwidth,trim={0cm 0cm 0cm 0cm},clip ]{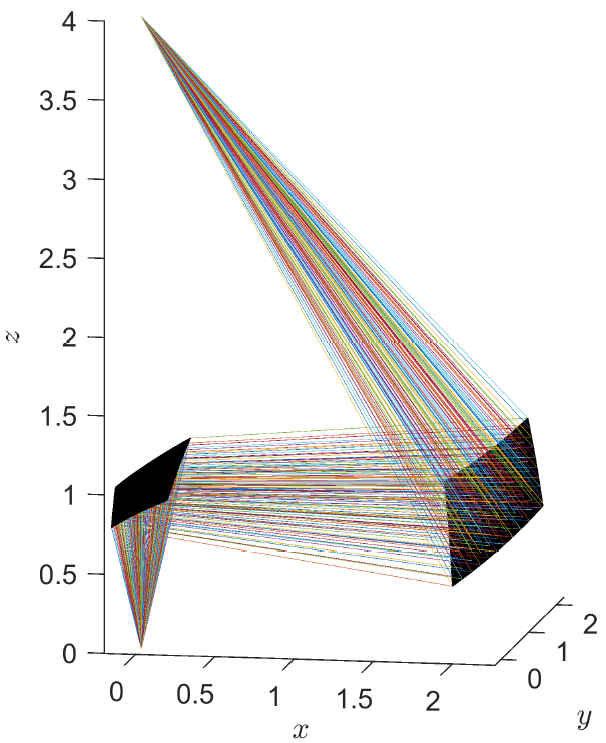}
    \caption{Reflector system with $h=1$.}
    \label{fig:3D_plot_2_1}
\end{subfigure}
\hfill
\begin{subfigure}[b]{0.43\textwidth}
    \includegraphics[width = 1.0\textwidth,trim={0cm 0cm 0cm 0cm},clip ]{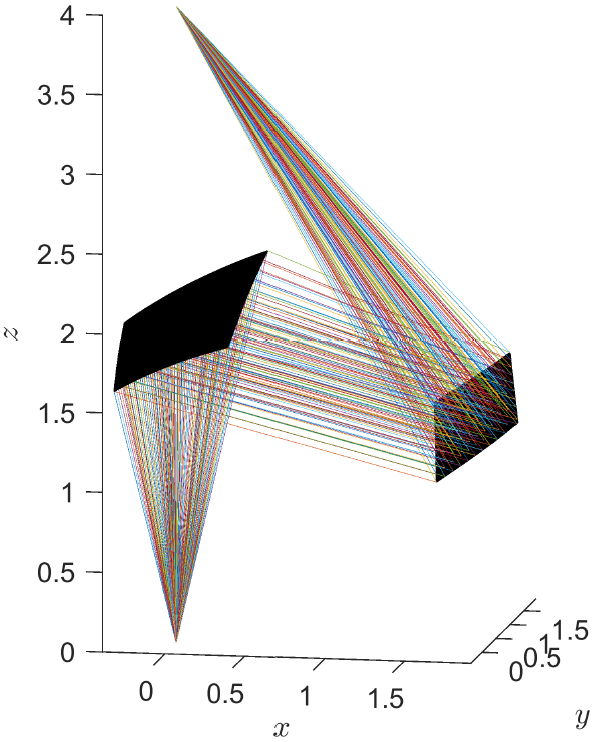}
    \caption{Reflector system with $h=2$.}
    \label{fig:3D_plot_2_2}
\end{subfigure}
\caption{Point-to-point two-reflector system with $V=8$, $\mathcal{S}=(0,0,0)$, $\mathcal{T}=(0,0,4)$, source and target domains given in Eqs. (\ref{eq:3D_domains_2}), $\alpha=10^{-2}$ after $100$ iterations on a $101\times101$ grid.}
\label{fig:3D_plot_2}
\end{figure}

Consider the point-to-point reflector system with point source $\mathcal{S}=(0,0,0)$ and point target $\mathcal{T}=(0,0,4)$ so that $\ell=4$. Furthermore, assume that the optical path length $V=8$ and $\alpha=10^{-2}$. Let
\begin{align}\label{eq:3D_domains_2}
    \bm{x}^T=(x_1,x_2)\in[-0.1,0.1]^2,&& \bm{y}^T=(y_1,y_2)\in[-0.3,-0.2]^2,
\end{align}
be the source and target domains, respectively. The source domain is uniformly discretized on a $101\times101$ grid. We also assume that the light at the source and target is distributed uniformly. Moreover, instead of taking the average distance of a light ray from $\mathcal{S}$ to $\mathcal{R}_1$, we can specify the distance $h$ from $\mathcal{S}$ to the center point $\bm{x}_c$ on $\mathcal{R}_1$ by rewriting Eq. (\ref{eq:transport_equation_stereographic}) as
\begin{subequations}
\begin{align}
    \left(v_1(\bm{x})+C\right)+\left(v_2(\bm{y})-C\right)=c(\bm{x},\bm{y}),
\end{align}
and specifically choosing
\begin{align}
    C=-v_1(\bm{x}_\text{c})+\log\left(\frac{(V^2-\ell^2)) (1+|\bm{x}_\text{c}|)}{2h\kappa_1(\bm{x}_\text{c})}-1\right),
\end{align}\end{subequations}
where $\bm{x}_\text{c}$ is the center point of the source domain. Here, the logarithmic term is obtained by substituting $\bm{x}=\bm{x}_\text{c}$ and $u_1=h$ in Eq. (\ref{eq:stereog_transport_u1_ptp_easy}). We choose $h=1$ and evaluate 100 iterations over scheme (\ref{eq:overview_iteration}), which gives us the reflector system in Fig. \ref{fig:3D_plot_2_1}. Since $\bm{x}_\text{c}=\bm{0}$, its corresponding ray of light reflects at the first reflector in the point $(0,0,1)$. If we choose $h=2$, we find the reflector system in Fig. \ref{fig:3D_plot_2_2}, where we see that the distance from $\mathcal{S}$ to the first reflector is indeed $2$ for the ray of light which corresponds with $\bm{x}_\text{c}=\bm{0}$. Therefore, this example shows that there are infinitely many reflector configurations and that a specific configuration can be chosen by fixing a specific point on the first reflector.

\begin{figure}[!ht]
  \centering
\begin{subfigure}[t]{0.48\textwidth}
    \vskip 0pt
    \centering
    \vspace{0.4cm}
    \includegraphics[width = 0.7\textwidth,trim={0cm 0cm 0cm 0cm},clip ]{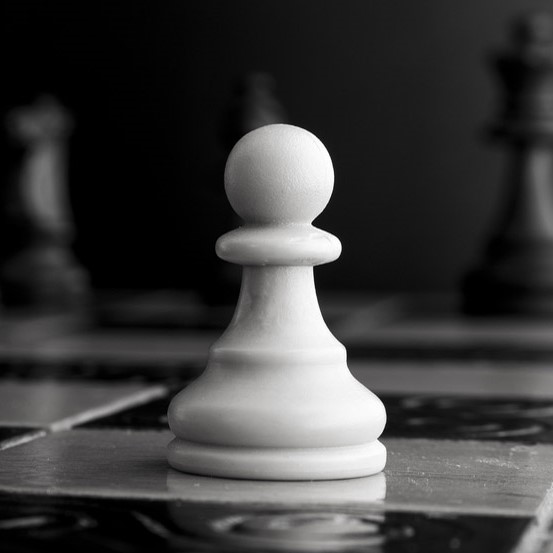}
    \vspace{0.4cm}
    \\\vspace{0cm}
    \caption{Image of a pawn \cite{PawnPicture}.}
    \label{fig:3D_plot_3_1}
\end{subfigure}
\hfill
\begin{subfigure}[t]{0.48\textwidth}
    \vskip 0pt
    \centering
    \includegraphics[width = 0.85\textwidth,trim={0cm 1.5cm 2.5cm 0.5cm},clip ]{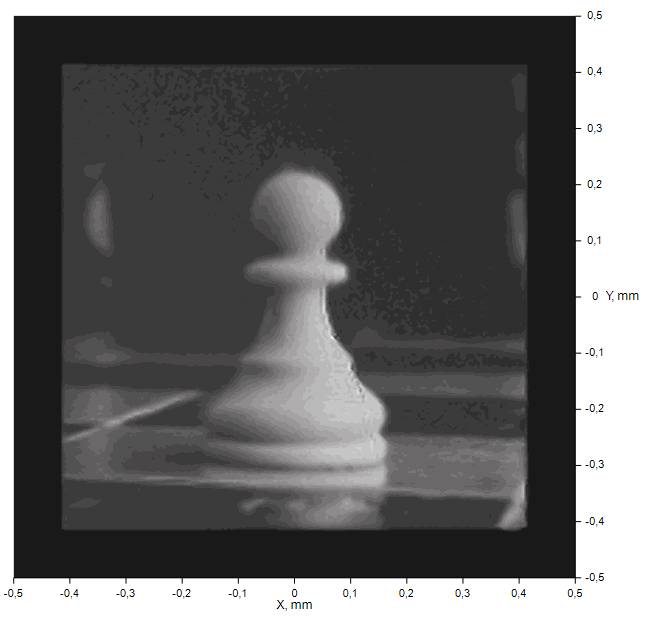}
    \caption{Illuminance pattern ray-tracing.}
    \label{fig:3D_plot_3_2}
\end{subfigure}
\\
\begin{subfigure}[b]{0.49\textwidth}
    \includegraphics[width = 1.0\textwidth,trim={0cm 0cm 0cm 0cm},clip ]{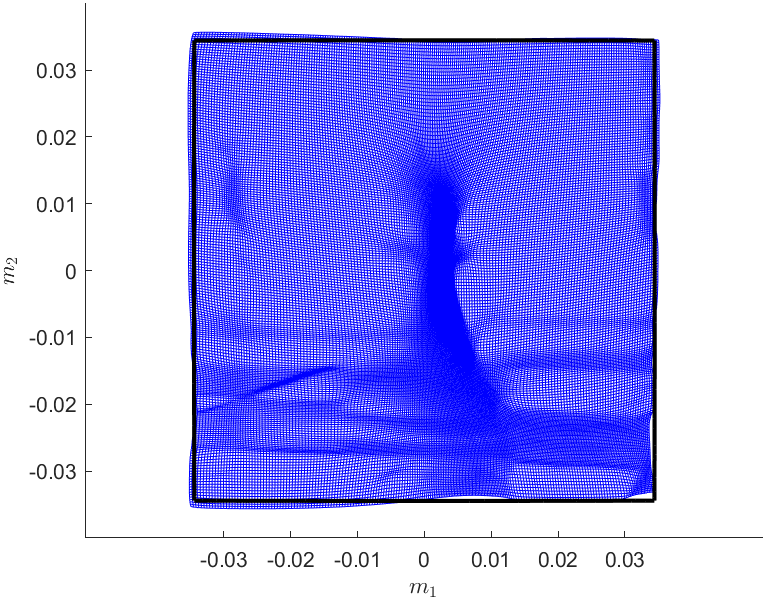}
    \caption{The mapping $\bm{m}$ after 5 least-squares iterations.}
    \label{fig:3D_plot_3_5}
\end{subfigure}
\hfill
\begin{subfigure}[b]{0.49\textwidth}
    \includegraphics[width = 1.0\textwidth,trim={0cm 0cm 0cm 0cm},clip ]{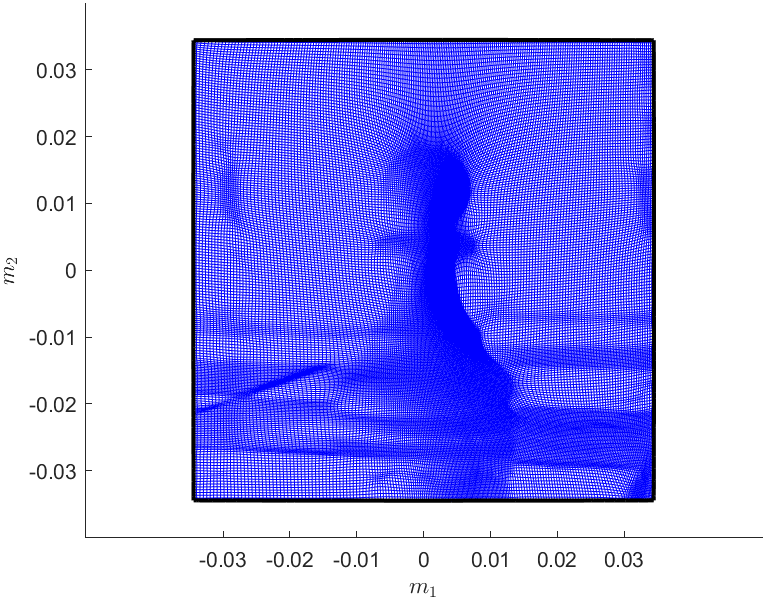}
    \caption{The mapping $\bm{m}$ after 100 least-squares iterations.}
    \label{fig:3D_plot_3_6}
\end{subfigure}
\\
\begin{subfigure}[b]{0.98\textwidth}
    \includegraphics[width = 1.0\textwidth,trim={0cm 0cm 0cm 0cm},clip ]{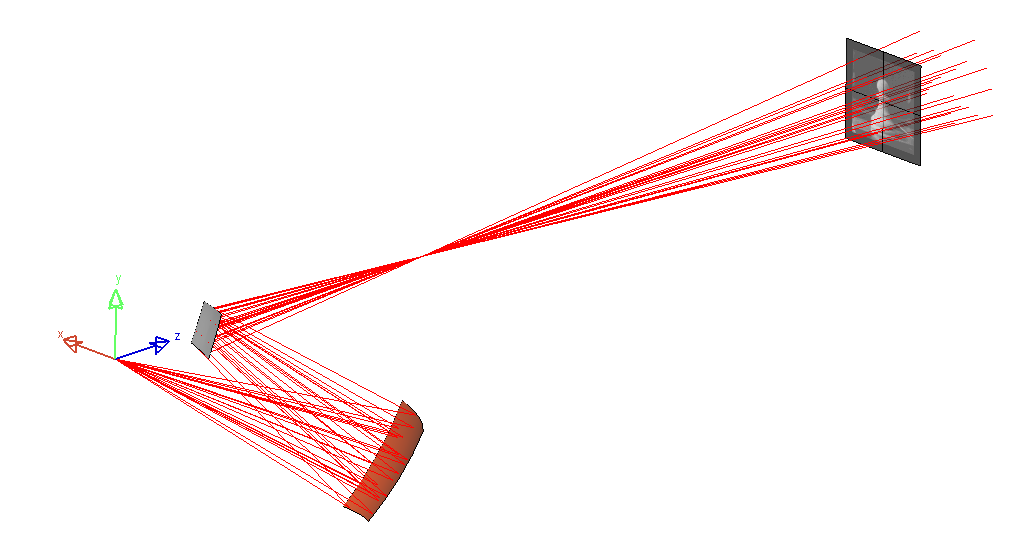}
    \caption{Ray-trace verification in LightTools.}
    \label{fig:3D_plot_3_3}
\end{subfigure}
\caption{Point-to-point reflector system with $V=8$, $\mathcal{S}=(0,0,0)$, $\mathcal{T}=(0,0,4)$, source and target domains given in Eqs. (\ref{eq:3D_domains_2}), $\alpha=10^{-2}$, $h=3$ and a $201\times201$ grid.}
\label{fig:3D_plot_3}
\end{figure}

\clearpage

\subsection{Mapping of a pawn}\label{sec:Pawn}
Consider the point-to-point two-reflector system with source $\mathcal{S}=(0,0,0)$ and point target $\mathcal{T}=(0,0,4)$ so that $\ell=4$. We also let $V=8$, $\alpha=10^{-2}$, $h=3$ and
\begin{align}\label{eq:3D_domains_3}
    \bm{x}^T=(x_1,x_2)\in[-0.5,-0.15]^2,&& \bm{y}^T=(y_1,y_2)\in[-0.035,0.035]^2,
\end{align}
are the source and target domains, respectively. The source domain is discretized on a $201\times201$ grid. We assume that light at the point source is distributed uniformly and that the target is distributed according to a gray-scale image of the pawn given in Fig. \ref{fig:3D_plot_3_1}.
\par
The least-squares algorithm first computes the mapping, which is after $5$ iterations illustrated in Fig. \ref{fig:3D_plot_3_5} and after $100$ iterations given in Fig. \ref{fig:3D_plot_3_6}. Clearly, the least-squares mapping converges to the desired image of the pawn in Fig. \ref{fig:3D_plot_3_1}. In the commercial ray-tracing software LightTools we verified the computed shapes of the reflectors. As a result, we obtained the reflector system in Fig. \ref{fig:3D_plot_3_3}, where $20$ light rays are shown and the light rays clearly converge to the point target after hitting both reflectors. Moreover, by placing the plane $z=10$ and keeping track of the rays hitting this plane, we find the ray-trace result obtained with $5\cdot10^7$ rays given in Fig. \ref{fig:3D_plot_3_2}. This illuminance pattern closely resembles the desired image of the pawn from Fig. \ref{fig:3D_plot_3_1}.

\section{Conclusion}\label{sec:conclusion}
In this paper we presented an inverse method for a reflector system where light originates from a point source, then propagates via two reflectors and finally ends up at a point target with a desired incoming light intensity. We introduced stereographic coordinates to describe the rays of light at the source and then used the generated Jacobian equation in combination with a least-squares algorithm to calculate the optical mapping and find the shapes of the reflector surfaces. Finally, we discussed how the height of a reflector can be fixed for a specific ray of light and presented numerical examples. Overall, the results advance our understanding of how the shape of the reflectors are obtained using an inverse method. Future research could use the least-squares algorithm to model other base optical systems \cite{Martijn1} and include physical phenomenon such as scattering effects \cite{Vi1}.

\appendix
\section{Sign check of the terms in Eq. (\ref{eq:transport_old})}\label{sec:appendix_A}
We show that the arguments of the logarithms in Eqs. (\ref{eq:transport_old}) are strictly positive. First of all, by the triangle inequality we have $b<d+u_2(\hat{\bm{t}})$ where $b$ denotes the distance between $P_1$ and $\mathcal{T}$. If we square both sides of this equation and on the left-hand side apply the cosine rule on triangle $\mathcal{S}P_1\mathcal{T}$, we see that
\begin{align*}
    u_1(\hat{\bm{s}})^2+\ell^2-2u_1(\hat{\bm{s}})\ell\hat{\bm{s}}\bm{\cdot}\hat{\bm{e}}=b^2<(d+u_2(\hat{\bm{t}}))^2.
\end{align*}
Next, by rewriting Eq. (\ref{eq:point_char_ptp}) to the form $d+u_2(\hat{\bm{t}})=V-u_1(\hat{\bm{s}})$ and substituting this in the above equation, we obtain
\begin{align*}
    0<\frac{V^2-\ell^2}{2Vu_1(\hat{\bm{s}})-2u_1(\hat{\bm{s}})\ell \hat{\bm{s}}\boldsymbol{\cdot}\hat{\bm{e}}}-1.
\end{align*}
Therefore, the argument of the logarithm in Eq. (\ref{eq:transport_u1_ptp}) is strictly positive. Similarly, we find that the argument of the logarithm in Eq. (\ref{eq:transport_u2_ptp}) is strictly positive. Consequently, we know that the left-hand side of Eq. (\ref{eq:big_transport_equation_ptp2}) is strictly positive, hence that the right-hand side is strictly positive and therefore that the argument in the logarithm of Eq. (\ref{eq:transport_c_ptp}) is strictly positive.

\clearpage

\begin{backmatter}
\bmsection{Funding}
Nederlandse Organisatie voor Wetenschappelijk Onderzoek (P15-36).

\bmsection{Disclosures}
The authors declare no conflicts of interest.

\bmsection{Data availability} Data underlying the results presented in this paper are not publicly available at this time but may be obtained from the authors upon reasonable request.
\end{backmatter}


\end{document}